\documentclass[runningheads]{llncs}
\usepackage{amsmath}
\usepackage{amsfonts}
\usepackage{graphicx} % Required for inserting images
\usepackage{url}
\usepackage{tikz}
\usepackage{subcaption}
\usepackage[braket, qm]{qcircuit}
\usetikzlibrary{chains,shapes.multipart}
\usetikzlibrary{shapes}
\usetikzlibrary{automata,positioning}
\usetikzlibrary{calc}
\usepackage{tikzscale}
\usepackage{makecell}

\title{Streaming IoT Data and the Quantum Edge: A Classic/Quantum Machine Learning Use Case}
\author{Sabrina Herbst, Vincenzo De Maio, Ivona Brandic}
\institute{Vienna University of Technology, Vienna, Austria,\\
\email{sabrina.herbst@tuwien.ac.at,\{vincenzo,ivona\}@ec.tuwien.ac.at}
}
\date{}
\authorrunning{Sabrina Herbst, Vincenzo De Maio, Ivona Brandic}
\titlerunning{The Quantum Edge: A Classical-Quantum ML Use Case}
\begin{document}

\maketitle

\begin{abstract}
With the advent of the Post-Moore era, the scientific community is faced with the challenge of addressing the demands of current data-intensive machine learning applications, which are the cornerstone of urgent analytics in distributed computing. Quantum machine learning could be a solution for the increasing demand of urgent analytics, providing potential theoretical speedups and increased space efficiency. However, challenges such as (1) the encoding of data from the classical to the quantum domain, (2) hyperparameter tuning, and (3) the integration of quantum hardware into a distributed computing continuum limit the adoption of quantum machine learning for urgent analytics. In this work, we investigate the use of Edge computing for the integration of quantum machine learning into a distributed computing continuum, identifying the main challenges and possible solutions. Furthermore, exploring the data encoding and hyperparameter tuning challenges, we present preliminary results for quantum machine learning analytics on an IoT scenario. 
\end{abstract}

\section{Introduction}
IoT data and machine learning have recently become the keystone of urgent computing~\cite{heidari2022machine,shafaf2019applications}. Data from IoT devices can be processed by machine learning models to improve simulations of different scientific phenomena~\cite{Lovholt2019}. However, machine learning applications require a huge amount of data for training. While data are streamed from IoT devices, they need to be transferred, stored and processed under strict response time constraints~\cite{demaio2022tarot}, which requires a huge amount of storage, computational and network resources. Therefore, training of machine learning  is often performed inside HPC facilities.

However, we recently entered the Post-Moore era, which faces the scientific community with the challenges of scaling computing facilities beyond current limits, which are codified by Moore's law and Dennard scaling. As a consequence, current HPC facilities struggle to scale with the increasing amount of data available, pushing the scientific community towards research in Post-Moore Computing to address this issue. Among different possibilities, quantum computing clearly stands out due to theoretical speedups and increased space efficiency. This is particularly true for quantum machine learning, whose potential benefits are (1) increased speed, (2) increased predictive performance, and (3) reduced amount of data needed for training~\cite{mensa2023quantum}.

Quantum machine learning requires adapting data from the classical to the quantum domain before training and inference, following a process that is often referred to as \emph{data encoding}. The choice of data encoding method can significantly affect the performance and accuracy of a quantum machine learning model~\cite{weigold2020data}. Also, the choice of hyperparameters is of capital importance for the performance and accuracy of trained models.

In this work, we expand our idea of the Quantum Edge~\cite{demaio2022} by investigating the possibilities of applying Edge computing methodologies to enable fast and efficient quantum machine learning on hybrid systems. After defining the problem, we present our idea of the Quantum Edge and describe how it can be applied to the target scenario. We identify challenges and possible solutions, and provide some preliminary results of our work towards the goal of Quantum Edge, tackling data encoding and hyperparameter selection.

We perform training and inference of quantum machine learning models based on a bike sharing dataset, publicly available at the UCI Machine Learning repository\footnote{\url{https://archive.ics.uci.edu/ml/datasets/Seoul+Bike+Sharing+Demand}. Accessed: 21.03.2023} \cite{dua_uci_2017}, which is representative of typical IoT data. We use IBM Qiskit~\cite{Qiskit} and Aer simulators to perform training and inference of quantum machine learning models, evaluating the predictive performance and runtime of different hyperparameters configurations.

The paper is organized as follows: first, we introduce preliminary notions about quantum computing and machine learning in Section~\ref{sec:background}. Related work is described in Section~\ref{sec:related}. Afterwards, we describe our motivational scenario in Section~\ref{sec:scenario}, and identify challenges of the target scenario in Section~\ref{sec:challenges}. In Section~\ref{sec:ideas}, we discuss possible solutions, while in Section~\ref{sec:results} we provide some preliminary results of our work, and we conclude our paper in Section~\ref{sec:conclusions}.

\section{Background}
\label{sec:background}
\subsection{Quantum Computing}
Quantum computing relies on different quantum phenomena, such as \emph{superposition}, \emph{entanglement}, and \emph{quantum parallelism}. The basic entities of quantum computing are quantum-bits, also known as \emph{qubits}. While classic bits can only be in two states, i.e., either $0$ or $1$, a single qubit $\psi$ is in a linear \emph{superposition} of the two \emph{orthonormal basis} states, $\ket{0}=[1,0]^{T}$  and $\ket{1} =[0,1]^T$, namely, $\ket{\psi} = c_0 \ket{0} + c_1 \ket{1}$, where $c_0$ and $c_1$ are the \emph{probability amplitudes}, which determine the probability that $\ket{\psi}$ will collapse in $\ket{0}$ or $\ket{1}$ when measured. We define $c_0, c_1 \in \mathbb{C}$, and $|c_0|^2 + |c_1|^2 = 1$. 

A set of $n$ qubits taken together forms a \emph{quantum register}. Quantum computation is performed by manipulating qubits within a quantum register. As a consequence, while a $n$-bit classic register can store one of $2^{n}$ values, a $n-$qubit quantum register can store $2^n$ values at the same time, resulting in a higher space efficiency. When measured, the quantum register will collapse into one single value $i$. The probability of collapsing into state $i$ depends on the value of the complex amplitude $c_i$. As a consequence, quantum computations are repeated several times, and the most frequent result will be the returned as the final result.

There are different models for quantum computation. In this work, we refer to the \emph{quantum circuit model}, where computation consists of the manipulation of qubits by applying a set of \emph{quantum gates}, similar to classical logical gates, and \emph{measurement} operations. Users prepare a high-level description of a quantum circuit~\cite{Qiskit}, that is transpiled to be executed on target quantum hardware.

\subsection{Quantum Machine Learning}
\label{sec:qml}
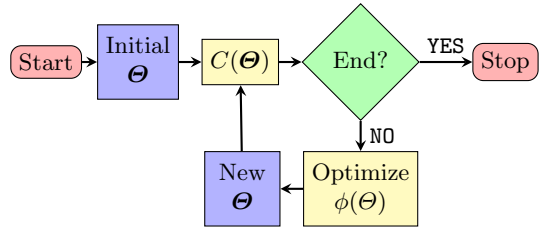
\begin{figure}[!h]
    \centering
    \tikzstyle{startstop} = [rectangle, rounded corners, minimum width=0.3cm, minimum height=0.3cm,text centered, draw=black, fill=red!30]
    \tikzstyle{process} = [rectangle, minimum width=1cm, minimum height=0.3cm, text centered, draw=black, fill=blue!30]
    \tikzstyle{decision} = [diamond, minimum width=0.3cm, minimum height=0.3cm, text centered, draw=black, fill=green!30]
    \tikzstyle{quantum} = [rectangle, minimum width=1cm, minimum height=0.3cm, text centered, draw=black, fill=yellow!30]
    \tikzstyle{arrow} = [thick,->,>=stealth]
    \begin{tikzpicture}
    \node (start) [startstop] {Start};
    \node (initial) [process, right=.2cm of start] {\makecell[c]{Initial\\$\vec{\Theta}$}};
    \node (quantum) [quantum, right=.3cm of initial] {\makecell[c]{$C(\vec{\Theta})$}};
    \node (decision) [decision, right=.3cm of quantum] {\makecell[c]{End?}};
    \node (c-theta) [quantum, below=.4cm of decision] {\makecell[c]{Optimize\\$\phi(\Theta)$}};
    \node (new-theta) [process, left=.3cm of c-theta] {\makecell[c]{New\\$\vec{\Theta}$}};
    \node (stop) [startstop, right=.7cm of decision] {Stop};
    \draw[arrow] (start) -- (initial);
    \draw[arrow] (initial) -- (quantum);
    \draw[arrow] (quantum) -- (decision);
    \draw[arrow] (decision) -- node[anchor=west]{\texttt{NO}} (c-theta);
    \draw[arrow] (c-theta) -- (new-theta);
    \draw[arrow] (new-theta) -- (quantum);
    \draw[arrow] (decision) -- node[anchor=south]{\texttt{YES}} (stop);
    \end{tikzpicture}
    \caption{Variational Quantum Algorithms.}
    \label{fig:vqas}
\end{figure}
Quantum machine learning (QML) aims at increasing time and space efficiency of machine learning tasks~\cite{biamonte2017quantum} by exploiting properties of quantum computing. We identify two main areas of QML: Quantum Classification and Quantum Regression. In this work, we focus on quantum regression.

We focus on Variational Quantum Algorithms (VQAs)-based regression, whose execution is summarized in Figure~\ref{fig:vqas}. The idea behind VQAs is to minimize a cost function $\phi$ that represents a specific property of a physical system (e.g., its ground state). The state of the physical system is modeled by a Parametrized Quantum Circuit $C$, which is a quantum circuit whose state is determined by a set of parameters $\vec{\Theta}$. The main idea is to minimize $\phi(\vec{\Theta})$ to achieve the optimal set of parameters $\vec{\Theta}^*$, which models the solution to our target problem.

\section{Related Work}
\label{sec:related}
First applications of hybrid classical/quantum systems are described by~\cite{Schulz2021}, focusing on the integration of quantum computers in HPC infrastructures. Similarly,~\cite{demaio2022} provides a proof of concept for hybrid molecular dynamics workflows. Applications of Edge computing for the integration of Non-Von Neumann architectures in the computing continuum have been discussed in~\cite{demaio2022escience}. VQAs are surveyed in~\cite{cerezo2021variational}.

The authors in \cite{oquinn_quantum_2020} provide a great overview of recent advances in QML, such as a hybrid implementation of the perceptron algorithm on current quantum hardware in \cite{tacchino_artificial_2019} which achieves an exponential advantage in storage. Furthermore, a quantum support vector machine was implemented in \cite{rebentrost_quantum_2014}, which scales logarithmically in terms of feature dimension and data samples.

In \cite{Srinivasan2017LearningHQ}, hidden quantum Markov models are investigated, showing that they manage to model the same classical data but with less hidden states.

Abbas et al. \cite{abbas_power_2021} investigate the capacity of VQAs for ML and compare the capacity of classical and quantum models, showing that some VQAs have desirable characteristics, allowing for a better expressivity and trainability, especially if the data encoding employed is difficult to simulate on classical hardware. 

Finally, the results from \cite{Srinivasan2017LearningHQ} show how QML can provide valuable insights both for quantum and classical computing \cite{oquinn_quantum_2020}.

\section{Motivational Use Case: Quantum Edge Analytics}
\label{sec:scenario}
\begin{figure}[!h]
\centering
\includegraphics[width=.6\columnwidth]{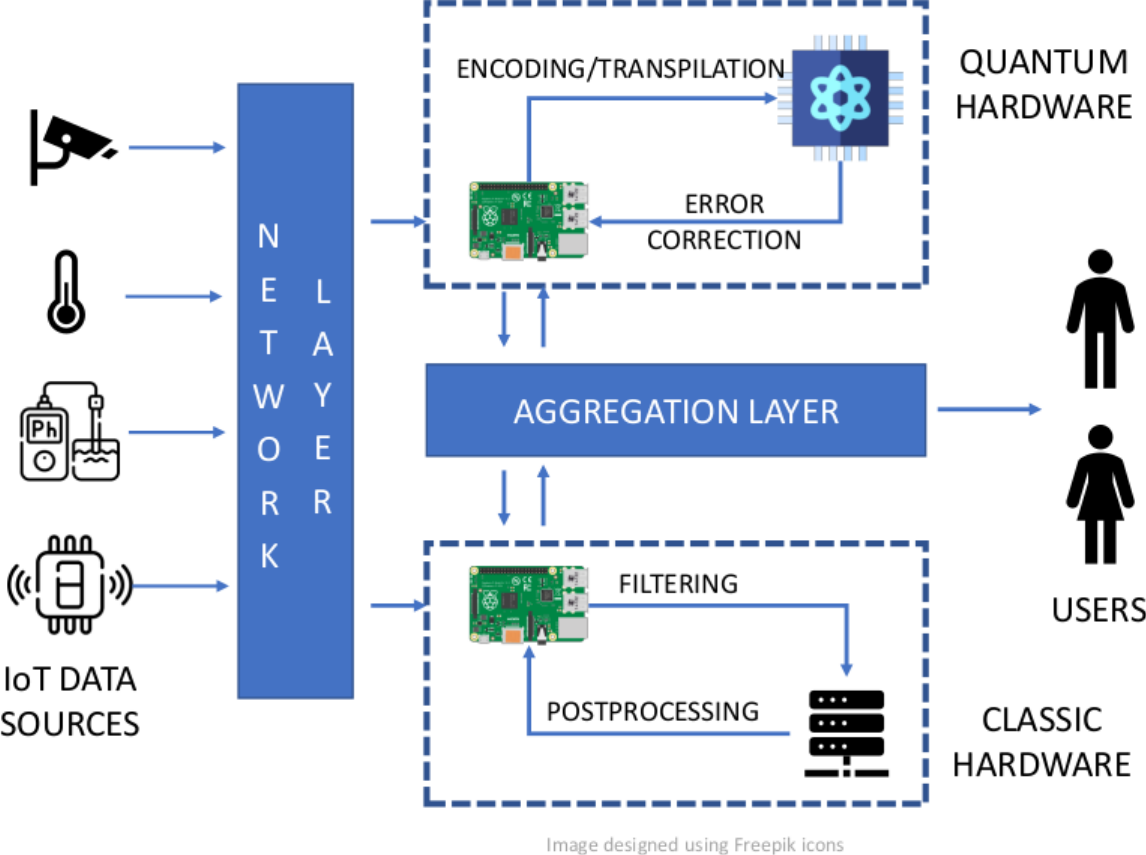}
\caption{The Quantum Edge Analytics.}
\label{fig:quantum-edge}
\end{figure}
Following from Section~\ref{sec:qml}, we identify three main hyperparameters in quantum regression: the \emph{encoding} of the input into a quantum state, the structure of the \emph{quantum circuit}, encoding the problem, and the \emph{optimizer} that is applied to identify the optimal set of parameters. Moreover, efficient execution of VQAs requires fast communication between classic and quantum systems, and eventually offloading some parts of the computation (i.e., optimization and encoding) for load balancing. We consider Edge computing to address these issues. 

Edge computing is a paradigm where data are processed in lower layers of the network, i.e., closer to data sources, such as sensors or edge nodes. Processing and storage at the Edge results in lower response time and enhanced data processing, which makes edge analytics suitable for near real-time analytics, such as traffic safety applications~\cite{lujic2021}, environmental monitoring~\cite{aral2022}, smart buildings~\cite{lujic2020}, and scientific computing~\cite{demaio2020}. Edge Computing will constitute an additional layer between classical and quantum machines, performing necessary tasks such as (1) encoding of streaming data, (2) performing small-scale simulations to reduce the load on quantum machines, (3) the transpilation of quantum circuits for the target quantum machine, or (4) the execution of classic optimizer. 

Figure~\ref{fig:quantum-edge} summarizes the concept of Quantum Edge Analytics. First, data coming from different IoT sources are streamed through the network to the Edge. At the Edge, resources of Edge nodes can be used (1) for encoding data into a quantum state, (2) for the transpilation of the quantum circuit, and (3) to apply error correction to the output of the computation. Also, Edge nodes can act as an aggregation layer between the quantum and classic layer. In the next sections, we describe the challenges of enabling our vision.

\section{Challenges}
\label{sec:challenges}

\subsection{Fast Data Encoding}
Data encoding is the process of translating classical data into a quantum state. Different methods are available in the literature~\cite{weigold2020data}. In QML, the most popular encoding is Feature Map Based Encoding~\cite{havlicek_supervised_2019,goto_universal_2021}, which is also what we consider in this work. In Feature Map Based Encoding, classical data are encoded employing a quantum feature map, i.e., a unitary function that maps the input data into a quantum feature space, where feature vectors are quantum states.

Data encoding is a computationally expensive process, that requires deep knowledge of (1) the input problem, (2) the target hardware where data will be processed, and (3) the algorithm employed for data processing. Data encoding is one of the keys for an efficient quantum computation, since inefficient data encoding could hinder speedups provided by QML. 

\subsection{Automatic Hyperparameter Tuning}
QML models, and in general VQAs, are influenced by different hyperparameter settings, such as (1) the structure of the parametrized quantum circuit (which includes hyperparameters such as the qubit entanglement and the circuit repetitions), (2) the classical optimizer, (3) the termination condition, (4) the cost function, and others~\cite{cerezo2021variational}. As shown by~\cite{demaio2022}, hyperparameter selection strongly affects both accuracy and the runtime of VQAs. 

The main issue with the selection of hyperparameters is that they strongly depend on the input problem and the target quantum machine, which makes it difficult to select them a-priori. Also, the high heterogeneity of quantum hardware, together with the wide spectrum of applications requiring Edge analytics, makes finding a "one-size-fits-all" solution infeasible. As a consequence, to enable Quantum Edge Analytics, it is of paramount importance to design methods for the automatic selection of hyperparameters, based on the available quantum hardware and target applications. 

\subsection{Edge Error Mitigation}
Current quantum computing hardware is referred to as Noisy Intermediate Scale Quantum (NISQ)~\cite{preskill2018quantum} technology, i.e., they are sensitive to the environment in which they are deployed and subject to quantum decoherence. As a consequence, different error correction methods need to be applied to the output of the quantum execution, while other approaches focus on exploiting noise to improve the performance of QML models~\cite{domingo2023taking}. Since Edge nodes are supposed to act as an aggregation layer between classical and quantum nodes, the application of error mitigation methods at the Edge is necessary to enable efficient data exchange. 

However, typical quantum error mitigation methods rely on computationally intense artificial neural networks~\cite{Kim2020,Bennewitz2022} and deep learning approaches~\cite{Kim2022}, which cannot be executed on typical Edge devices, i.e., Raspberry Pi~\cite{steffenel2021assessing}.

\section{Quantum Edge Pre/Post Processing}
\label{sec:ideas}
\subsection{Fast Data Encoding}
Considering the high rate at which IoT data are streamed over the network~\cite{demaio2022tarot}, enabling Quantum Edge analytics requires that data are encoded at a speed that allows data to be quickly transferred to the quantum hardware to be encoded as a quantum state. Typical data encoding methods~\cite{weigold2020data} require performing complex algebraic operations, such as computing unitary functions and complex matrix operations, which is computationally intensive for typical Edge nodes.

To enable fast data encoding, our suggested solution is to increase heterogeneity at the Edge, by enabling the use of a wider spectrum of accelerators that can perform data encoding at a higher speed, i.e., GPUs, FPGAs or ASICs~\cite{hu2022survey} such as Nvidia Jetson Nano, or Edge Spartan 6, which are already available off-the-shelf. Another possibility is to exploit Edge preprocessing to apply different techniques, such as symbolic data representation, to reduce the size of data that needs to be forwarded to quantum nodes, or enable distributed data encoding at the Edge, exploiting multiple Edge devices to perform the required calculations.

\subsection{Automatic Hyperparameter Tuning}
Hyperparameter tuning can affect both the predictive performance and runtime of QML models, and it is inherently correlated to (1) the input data coming from IoT devices, (2) the target problem, and (3) the available quantum hardware. Since Edge nodes act as an intermediate layer between data sources and classic/quantum hardware, they have access to this information. Therefore, it must be their responsibility to determine the ideal setting of hyperparameters to perform QML, with minimum or no human intervention in the process. 

Our suggestion is to enable Edge nodes to perform hyperparameter tuning techniques, such as Bayesian Optimization or Grid Search~\cite{wu2023hyperparameter}, based on data available from previous executions of QML models. Interplay between Cloud and Edge resources, as it is common in Edge AI~\cite{cao2022decentralized}, represents another possibility to cope with the limited resources available at the Edge.

\subsection{Edge Error Mitigation}
The output of QML inference must be postprocessed at the Edge before being forwarded to users due to the amount of noise in modern NISQ hardware. However, typical techniques for error mitigation rely on complex ML models, which are not suitable for execution at the Edge. 

Enhancing Edge devices with specific accelerators, such as TPUs or FPGAs allow us to speed up the execution of error correction methods. Another promising techniques is to  downsize ML models, i.e., running a reduced size model for error correction, optimized for specific quantum hardware, while training is performed at the HPC layer. %

\section{Preliminary Results}
\label{sec:results}

In this section, we perform first preliminary evaluation on Quantum Regression of IoT streaming data. We focus on Feature Map based encoding, and hyperparameter configurations. We use the Seoul Bike Sharing Dataset\cite{sathishkumar_using_2020,sathishkumar_rule_2020}, which is available at the UCI Machine Learning Repository\footnote{\url{https://archive.ics.uci.edu/ml/datasets/Seoul+Bike+Sharing+Demand}. Accessed: 21.03.2023}~\cite{dua_uci_2017}, since its features include data similar to those in IoT sensors (i.e., temperature, humidity, wind speed).

\subsection{Experimental Setup}
We use the Python programming language and the open source package Qiskit \cite{Qiskit} for our experiments. We employ Quantum simulators from Qiskit Aer, which provide the possibility of simulating 'perfect' quantum computers, without the noise that is present in current hardware.

Our dataset has $13$ attributes, an integer target and $8,760$ instances. The goal is to predict bike sharing demands for every hour of the day, based on weather conditions and seasonal information. We normalize the target into the range $[0,1]$, as Qiskit's VQR function does not allow predicting bigger values.

We perform an exhaustive search over all hyperparameter configurations, summing up to 672 configurations. We use principal component analysis (PCA) to reduce the dimensionality to seven. This was done to ensure that the experiments could be run on IBM quantum computers as well, as they offer seven qubit machines for free. %

We limited our experiments to 400 data points for training and 250 for evaluation to ensure a reasonable runtime. %
Table~\ref{tab:top_five} summarizes the top five configurations in terms of Mean Square Error (MSE), Mean Absolute Error (MAE) and running time. More configurations are described in the following sections.

We conducted experiments on a Debian/GNU Linux 11 computer, featuring an Intel(R) Xeon(R) CPU E5-2623 v4 (16 cores @ 2.60GHz) and 128 GB RAM.

\begin{table}[]
    \centering
    \begin{tabular}{|c|c|c|c|c|c|c|}\hline
        Ansatz & Optimizer & Feature Map & Entanglement & MSE & MAE & Time \\\hline
        EfficientSU2 & SPSA & ZFeatureMap & circular & 0.0209 & 0.1074 & 23993s \\
        RealAmplitudes & SPSA & ZFeatureMap & full & 0.0219 & 0.1113 & 27585s \\
        TwoLocal & SPSA & ZFeatureMap & sca & 0.0227 & 0.1172 & 22055s \\
        EfficientSU2 & SPSA & ZFeatureMap & full & 0.0229 & 0.1161 & 31995s \\
        TwoLocal & SPSA & ZFeatureMap & linear & 0.0245 & 0.1209 & 21314s \\\hline
    \end{tabular}
    \caption{Top Five Configurations}
    \label{tab:top_five}
\end{table}

\subsection{Evaluation}

\subsubsection{Data Encoding}
\begin{figure}[!h]
    \centering
    \includegraphics[width=.5\columnwidth]{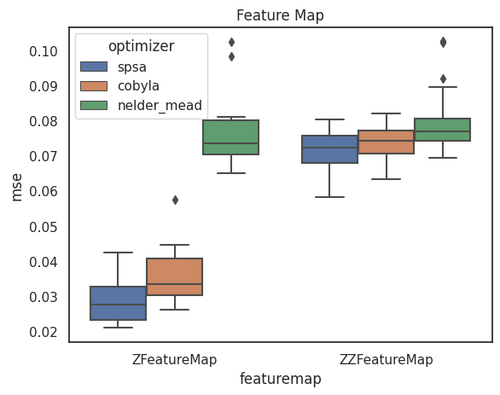}
    \caption{MSE of Training w.r.t. Feature Map}
    \label{fig:data-encoding}
\end{figure}

Figure~\ref{fig:data-encoding} summarizes the results of data encoding. %
Data encoding defines how classical data is mapped into a quantum state. An example is \emph{Basis Encoding}, where a number, i.e. $5$, is transformed into binary $101$ and the bits are directly mapped into qubits $\ket{101}$. We employ a more space efficient feature map, that is estimated to be difficult to reproduce on a classical computer, i.e. the PauliFeatureMap\cite{havlicek_supervised_2019}. It uses a feature map to map the data into a quantum state and two concrete implementations are derived in Qiskit, i.e., ZFeatureMap and the ZZFeatureMap. Since a qubit entanglement strategy can be specified for ZZFeatureMap, we experiment with 'full' entanglement (all qubits are entangled with all other ones), 'linear' ($q_i$ with $q_{i+1}, i \in [0, N-2]$, N being the number of qubits), 'circular' (linear plus $q_{N-1}$ with $q_0$), 'pairwise' (in even layers, qubit $q_i$ is entangled with $q_{i+1}$, in uneven ones with $q_{i-1}$) and 'sca'. In 'sca' entanglement, the target and control qubit are swapped in every iteration, and the entanglement between qubit $q_{N-1}$ and $q_0$ is shifted in every iteration, i.e. in the first repetition it is the first entanglement operation, in the second it is shifted into the second position and $q_0$ and $q_1$ are entangled first.

Figure \ref{fig:data-encoding} shows the performances of different feature maps, highlighting the optimizers used. It can be seen that ZFeatureMap works significantly better than ZZFeatureMap does, however the advantage is not independent of the optimizer. Our results show that choosing the feature map carefully can significantly impact the results. Nonetheless, a good feature map cannot make up for a bad optimizer.

\subsubsection{Circuit Hyperparameter Tuning}
\begin{figure}
    \centering
    \begin{subfigure}{.48\columnwidth}
        \includegraphics[width=\textwidth]{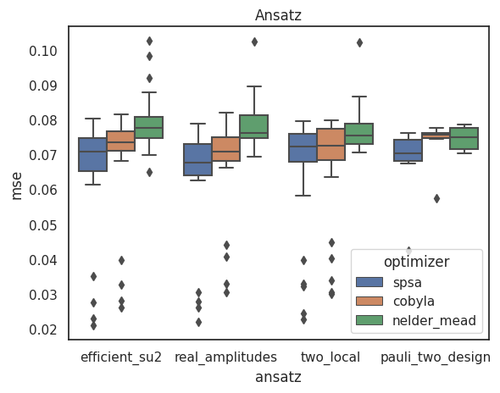}
        \caption{MSE w.r.t. Ansatz Configuration.}
        \label{fig:res-ansatz}
    \end{subfigure}
    \begin{subfigure}{.48\columnwidth}
        \includegraphics[width=\textwidth]{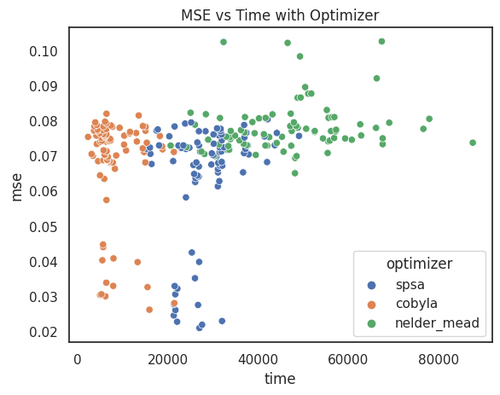}
        \caption{MSE w.r.t. Optimizers.}
        \label{fig:res-optimizer}
    \end{subfigure}
    \label{fig:hyp-tuning}
\end{figure}

Figure~\ref{fig:res-ansatz} summarizes the results for the Ansatz selection. EfficientSU2, TwoLocal, RealAmplitudes and PauliTwoDesign are commonly used ansatzes available in Qiskit. They all use rotations around the X, Y and Z axis of the Bloch sphere, which is a way to geometrically represent a qubit, with a trainable angle $\theta$. Furthermore, entanglement strategies can be employed for all except PauliTwoDesign, analogously to the ZZFeatureMap.

Figure~\ref{fig:res-optimizer} summarizes results for selection of different optimizers. We select COBYLA \cite{powell_direct_1994}, SPSA \cite{spall_overview_1998}, and Nelder-Mead \cite{nelder_simplex_1965}, because they have diverse characteristics, which allows to gain insights on which class of optimizers could work well for VQAs. COBYLA and Nelder-Mead are gradient-free methods which use trust regions and the simplex algorithm, respectively, to find an optimization path. SPSA calculates the loss function using only two measurements and is often recommended in noisy environments\footnote{\url{https://qiskit.org/documentation/stubs/qiskit.algorithms.optimizers.SPSA.html. Accessed 17.06.2023}}. %
We can see that the ansatzes perform similarly in mean, however, for all except PauliTwoDesign, some configurations that perform a lot better can be found. Furthermore, the range that the PauliTwoDesign configurations form is a lot more narrow than for the other ansatzes.

When analyzing the optimizers, we see that SPSA makes up the best configurations, however, there are several good COBYLA ones as well. Despite taking a long time to optimize, Nelder-Mead is significantly outperformed by all other optimizers. Furthermore, the SPSA runtime is usually higher than COBYLA, which makes COBYLA better in case of runtime constraints. 

\section{Conclusion and Future Work}
\label{sec:conclusions}
In this work, we investigate the possibility of applying QML on streaming IoT data. First, we describe the concept of the Quantum Edge, i.e., applying Edge computing as an integration layer between classic and quantum hardware. We describe challenges of enabling our vision and identify possible solutions. Finally, we provide some preliminary results on the feasibility of applying QML for streaming IoT data. 

In the future, we plan to investigate applications of QML to different application scenarios, which will help data scientists in the application of QML on IoT data. Also, we will further investigate the Quantum Edge, addressing challenges of integration of classic and quantum hardware by means of Edge computing.

\section*{Acknowledgements}

This work has been partially funded through the Rucon project (Runtime Control in Multi Clouds), Austrian Science Fund (FWF): Y904-N31 START-Programm 2015, by the CHIST-ERA grant CHIST-ERA-19-CES-005, Austrian Science Fund (FWF), Standalone Project Transprecise Edge Computing (Triton), Austrian Science Fund (FWF): P 36870-N, and by Flagship Project HPQC (High Performance Integrated Quantum Computing) \# 897481 Austrian Research Promotion Agency (FFG). We acknowledge the use of IBM Quantum services for this work. The views expressed are those of the authors, and do not reflect the official policy or position of IBM or the IBM Quantum team.

\bibliographystyle{unsrt}
\bibliography{quickpar-invited}

\end{document}